# Suspicious electron-doped CMR in sintered La$_{1-x}$Ce$_x$MnO$_3$

Hsiung Chou*, C. B. Wu and S. G. Hsu

*Department of Physics and Center for Nanoscience and Nanotechnology, National Sun*

*Yat-Sen University, Kaohsiung 804 , Taiwan*

## Abstract

La$_{0.7}$Ce$_{0.3}$MnO$_3$ (LCeMO) is known usually called an electron doped colossal magnetotesistance material, which contains multiphase of n-type LCeMO, CeO$_2$ and Mn-O. However, the present energy dispersion spectroscopy measurement shows that the La$_{0.85}$Ce$_{0.017}$MnO$_{3+\delta}$ is the main phase in the bulk solid of material, which doesn't have the LCeMO phase.  This finding is consistent with the XRD data analysis which indicates that the composition is La$_{0.87}$Ce$_{0.014}$MnO$_{2.93}$.  Since the highly localized $f$ electrons of Ce ions may not be transferred to Mn ions significant oxygen deficiency is needed to drive La$_{0.87}$Ce$_{0.014}$MnO$_{2.93}$ into n-type state that, which conflicts with the observation of a low T$_C$ fact of the vacuum annealed sample and to the natural tendency for the absorption of excess oxygen that yields a positive δ.  Similar properties are also found in La$_{1-x}$Te$_x$MnO$_{3+\delta}$, La$_{1-x}$Sn$_x$MnO$_{3+\delta}$ and La$_{1-x}$Zr$_x$MnO$_{3+\delta}$ systems.  The present study casts doubt on the interpretation of n-ype electronic property in bulk solid of LCeMO and proposed that the n-type CMR may be formed in metastable state.

Colossal magnetoresistance (CMR) materials have many interesting fundamental properties [1,2] and potential applications in information industry. By substituting x fraction of divalent alkaline-earth ions at the trivalent La sites of $LaMnO_3$ (LMO), i.e. $(La_{1-x}B_x)MnO_3$, an x portion of $Mn^{3+}$ ions was usually assumed to be converted into $Mn^{4+}$ ions, which renders insulated parent compound into ferromagnetic conductors [3-6]. The substitution of tetravalent Ce [7-21], Te [22-26], Sn [27-29] or Zr [30] ions at the La sites was assumed to yield a $Mn^{2+}$ and $Mn^{3+}$ mixed state. It was argued that the strong electron-phonon coupling [31] might cause electron doped $LaMnO_3$ to behave as hole doped. Based on these assumptions, p-n oxide junction or spintronic device might be realized in the future [32-35]. However, the single phase tetravalent doped CMR, such as $La_{0.7}Ce_{0.3}MnO_3$ (LCeMO), has not been realized in the bulk form [13]. LCeMO was reported to contain substantial $CeO_2$ [13] and Mn-O [8] impurites by powder x-ray diffraction (XRD) patterns and magnetization measurements. Ganguly, Gopalakrishnan and Yakhmi suspected that LCeMO has a lanthanum deficient $La_{0.7}MnO_3$ phase with Pbnm symmetry [10]. However, most follow up papers [8,12,13,19,25-28,30] still regarded these compounds as n-type LCeMO samples simply followed earlier reports [9,10,16,22-24,29]. The electronic states of the $La_{0.7}T_{0.3}MnO_3$ (T=tetravalence ions) have been studied by x-ray absorption spectroscopy (XAS) [14,15,17,24], Photoemission spectroscopy [11,20,24,25,29] and Iodometric titration method [9]. X-ray absorption spectroscopy (XAS) found that the Ce ion is in the 4+ state and the Mn ion has a valence between 2+ and 3+, and concluded that $La_{0.7}Ce_{0.3}MnO_3$ was possibly in the n-type state. $La_{0.7}Ce_{0.3}MnO_3$ was identified earlier by XRD measurement to a single phase by Mandel and Das [7] and Philip *et.al.* [9]. However, the same compound was found to contain $CeO_2$ impurities by Mitra *et.al.* later [13]. The Mn-O impurity was found by the low temperature antiferromagnetic magnetization measurements [8]. The ignorance of the presence of impurities might be the cause of misinterpretations of XAS data to conclude that the compound is n-type.

Hall measurements can identify the type of carrier. However, the Hall measurements of the *in situ* epitaxial film grown by Raychaudhuri *et al* [17] showed the compound to be n-type, while the post-annealed epitaxial films grown by Yanagida *et al* [19] found it to be p-type. Yanagida *et al* also observed segregation of the Ce-rich and Mn-deficient nanoclusters from a lanthanum deficient manganite, which was argued to be responsible for the ferromagnetic-metal to paramagnetic-insulator transition. Since Raychaudhuri *et.al.* did not specify the microstructure or composition of their samples, phase segregation could also present in their samples.

In this study, $La_{0.7}Ce_{0.3}MnO_3$ compound was formed by standard solid state reaction by mixing 99.995% pure $La_2O_3$, $CeO_2$ and $MnO_2$ powders. The initial mixture was calcined at $900^oC$ and $1050^oC$ in air for 24 hours. The powder was then pressed into pellets and sintered at $1280^oC$ in air for 72 hours with several intermediate grinding processes. The magnetization and transport properties and the XRD pattern were measured and compared with previous reports to ensure the formation of the similar compound. The magnetization and electric transport measurements were carried out by the supercoducting interference device (SQUID) and a standard four points probe capable of applying magnetic field up to 1 Tesla. The compositional distribution was taken quantitatively by energy dispersion spectrum (EDS) in a field emission scanning electron microscopy (SEM).

The sample was then investigated by EDS and SEM to find the presence of $CeO_2$ and Mn-O impurities. EDS excluded the presence of the n-type $La_{0.7}Ce_{0.3}MnO_3$ phase. EDS found only the La-deficient $La_{0.9}MnO_{3+\delta}$ (LMO) phase [36,37] with the low Ce-content $La_{0.85}Ce_{0.02}MnO_{3+\delta}$ as the primary phase. The XRD measurements had the same observation as those of EDS. The diffraction pattern, which was fitted to the n-type $La_{0.7}Ce_{0.3}MnO_3$ with a orthorhombic structure with the Pnma space group [8] is now found to be better fitted the Lanthanum- and oxygen-deficient $La_{0.87}Ce_{0.02}MnO_{2.93}$ with a rhombohedral structure with $R\bar{3}C$ space group. This finding suggests that the previous assignment of electronic valence of Ce and Mn ions might be due to a misinterpretation of contributions from impurities, which prompts a need to clarify whether the Ce doped $LaMnO_3$ compound is a n-type materials.

The magnetization and resistivity as functions of the temperature are plotted in Fig. 1. The R-T curve (solid curve) exhibits two metallic-like-insulating-like transition peaks at 266K and 240K. The zero-field-cooling (ZFC) remnant magnetization measurement indicates that a very sharp paramagnetic-ferromagnetic transition at ~266K, which essentially coincides with the R-T insulator-like to metallic-like transition at 266K. The antiferromagnetic transition at ~50K observed by Gebhardt, Roy and Ali, which was attributed to the $MnO_2$ compounds [8], is hard to be identified in the present study. The XRD characteristic peaks of the Mn-O compound are not observed in the current XRD data, as shown in the Fig. 2. Fig. 2 shows the characteristic XRD peaks of $CeO_2$ consistent with the observation of Mitra *et.al.*, though Mitra *et.al.* attributed them as evidence of n-type $La_{0.7}Ce_{0.3}MnO_3$. The $CeO_2$ implies the presence of $Ce^{4+}$, which may lead to misinterpretation as electron doped LCeMO.

In complementary to XRD measurements the energy dispersion spectroscopy (EDS) and a scanning electron microscopy (SEM) may provide better understanding of LCeMO. To calibrate the cation reading, large area images over mm scale were taken for $La_{0.7}Ca_{0.3}MnO_3$ and $LaMnO_{3+\delta}$ synthesized by the solid state reaction method. The samples have been characterized to have a single phase by XRD measurements. Figure 3 shows the back scattering electron image (BEI) of the bulk $La_{0.7}Ce_{0.3}MnO_3$. In BEI, heavier atoms (or ions) yield larger back scattering cross section and exhibit as white image, whereas the light atoms (ions) exhibit in dark image. The scattered white droplets are Ce-rich compounds for Ce ions are the heaviest ions in the samples. La and Mn ions can hardly be found even though the electron beam has a long penetration depth that can penetrate through the thin part of the irregular droplet. The black blocks are a few micrometers in size and contain light ions. EDS reading confirms that these black blocks are Mn-O compounds. However, the amount of the oxygen content is uncertain in the present EDS measurements. Nevertheless, the large grain size of the Mn-O compound contributes insignificant XRD features that can only be traced by the Rietveld simulation [35]. The major morphology of the sample is the huge gray grains with grain sizes ranged from a few micrometers to nearly 10 micrometers, which distribute over the most part of the figure. The EDS scanning indicates that the gray grains are basically $La_{0.9}MnO_{3+\delta}$ with very limited solubility of Ce to form $La_{0.85}Ce_{0.017}MnO_{3+\delta}$.

The main peaks in the XRD pattern that was proposed to be the LCeMO phase was very similar to those of $La_{0.894}MnO_3$ (LMO) [37,38] shown in Fig. 2. Both LCeMO and LMO phases can used to fit the x-ray data using the same X-ray analysis technique of Rietveld [36]. By fitting the XRD pattern with Mitra $et.al$'s parameters for the Pnma space group and the orthorhombic structure with $a$=5.508(7)Å, $b$=5.543(2)Å and $c$=7.833(5)Å and with the presence of $CeO_2$, the result shows a large residual (error) at the locations of all peaks with $\chi^2$=10.75. The peaks at around 31.5°, 37°, 40.5°, 58.5°, 68.5° and 78.5° in the current XRD in Fig. 3(a) disagree with fitted ones. The peak at around 26° predicted by the fit did not exist in XRD data. In Fig. 3(b), the structure parameters of Dezanneau's [37] (with the space group= $R\overline{3}C$, rhombohedral structure, and $a$= $b$=5.5097(1)Å, $c$=13.3395(3) Å and $\alpha=\beta=90°$, $\gamma=120°$) and of $CeO_2$ and $Mn_3O_4$compounds are used to fit the XRD data. The presently fitted structured parameters have higher accuracy of $\chi^2$=4.059 as shown in the Fig. 3(b). The two peaks, namely the tiny peaks at 31.5° right on the left prominent with maximum intensity and the small independent peak at 37° can not be fitted by this combination. Based on this result, one may assume that a limited amount of Ce may dissolve into the structure as $La_{0.9}Ce_\varepsilon MnO_{3+\delta}$, where ε is the solubility of Ce. By using the composition $La_{0.85}Ce_{0.017}MnO_3$ based on the EDS observation in combination with $CeO_2$ and Mn$_3$O$_4$ the fitted XRD pattern, Fig. 3(c), which has a higher accuracy ($\chi^2$=2.909) with main phase to be La$_{0.87}$Ce$_{0.014}$MnO$_{2.934}$ ($a$=$b$=5.531(8)Å, $c$=13.370(6) Å and $\alpha$=$\beta$=90$^o$, $\gamma$=120$^o$).

The combined content of La and Ce is about 0.88, which is similar to Dezanneau *et.al*'s finding of La$_{1-x}$MnO$_{3+\delta}$ with x>0.1 that is always decomposed into La$_{0.9}$MnO$_{3+\delta}$. The very small solubility of Ce does not contribute sufficient electron, so that a large oxygen deficiency, $\delta$≤-0.1, is needed to drive La$_{0.87}$Ce$_{0.014}$MnO$_{2.934}$ into n-type region, which conflicts with the fact of low T$_C$ of the air annealed samples [7] and to the natural tendency that prefers a positive $\delta$ [22].

The most direct way to investigate the type of carriers in thin films is the Hall measurement. Raychaudhuri [17] and Yanagida *et. al.* [19] have carried out the Hall measurements for epitaxial LCeMO films and Gao *et. al.* [29] for LSnMO films. The carriers in the *in-situ* epitaxial films grown by Raychaudhuri were concluded to be conduction electrons only for T<T$_C$ by assuming a simple spherical Fermi surface. When the expitaxial films were fabricated at high temperature with post annealing [19], Ce-rich nanoclusters were formed to precipitate from the matrix leaving a lanthanum deficient manganite which is responsible for the magnetic and electric transition with p-type carriers. The growth of thin films is usually a non thermal equilibrium process. The *in-situ* expitaxial film can form a metastable phase. In contrast, the fabrication process used in this study is driven by thermal energy under thermal equilibrium, which yielded a p-type La$_{0.87}$Ce$_{0.017}$MnO$_{2.93}$.

Similar dependence on the fabrication technique has been observed in La$_{1-x}$Te$_x$MnO$_{3+\delta}$

(LTeMO) [22-26], $La_{1-x}Sn_xMnO_{3+\delta}$ (LSnMO) [27-29] and $La_{1-x}Zr_xMnO_{3+\delta}$ (LZrMO) [30] systems. In LSnMO and LZrMO systems, the coexistance of $Mn_2O_3$, $Mn_3O_4$, $LaMnO_{3+\delta}$ or $La_2Sn_2O_7$ impurities had been confirmed by XRD [28]. Even in the LTeMO system, which exhibited the best resolved XRD pattern, the EDS investigation indicates that the final compound by pre-treated in a oxygen free atmosphere, the Ar atmosphere, contains very limited amount of Te can hardly been detected by EDS. The Te might be evaporated during Ar annealing processes and forming very few $TeO_2$. The phases in the LTeMO system are found to be $La_{0.9}MnO_{3-\delta}$, Mn-O and Te-O. These impurity phases provide signals in XAS and photoemission spectroscopy measurements that were misinterpreted as the evidence of n-type CMR. The $T_C$ of LTeMO system increases with the doping of Te and with the amount of oxygen content for $x \leq 0.15$, which implies the increasing of $T_C$ might be attributed to the hole doping due to the excess oxygen content [22]. Therefore, the investigation of microstructure by SEM and EDS in these systems is essential.

To conclude, we have examined the microscopic morphology and composition distribution of $La_{0.7}Ce_{0.3}MnO_3$ bulk by a SEM and EDS techniques. EDS excludes the presence of the $La_{0.7}Ce_{0.3}MnO_3$ phase, instead, a La deficient $La_{0.87}Ce_{0.014}MnO_{2.934}$ is observed. This finding gets mutual support by reanalyzing the XRD pattern which has been misinterpreted by other groups. Similar situations might have happened in LTeMO, LSnMO and LZrMO systems which prompts a need to clarify whether the Ce doped $LaMnO_3$ compound is a n-type materials. From the LCeMO films grown *in-situ* and by post annealed reveal that the n-type CMR may be formed in a metastable state but not in the thermal equilibrium state.

We would like to acknowledge our great appreciation to Professors M.-H. Tsai, H. D. Yang and S. J. Sun for their precious discussion and suggestions, and to Mr. C. L. Huang for the assistance in running the Rietveld program. This project is supported by National Science Council in Taiwan under project NSC 94-2112-M-110 -004.

014426 (2003).

**Captions:**

FIG. 1: The resistivity and the zero field remnant magnetization of $La_{0.7}Ce_{0.3}MnO_3$ as a function of temperature.

FIG. 2: The comparison of x-ray diffraction patterns of Mitra's result, (a), of the present data, (b); and of $La_{0.9}MnO_{3+\delta}$, (c).

FIG. 3: The Rietved refinement results for using (a) Mitra's result, the orthorhombic structure with Pnma space group and $a$=5.5087Å, $b$=5.5431Å and $c$=7.8335Å; (b) Dezanneau's result, the rhombohedral structure with $R\bar{3}C$ space group, and the $CeO_2$ impurity; (c) EDS result of $La_{0.85}Ce_{0.02}MnO_{3+\delta}$ and $CeO_2$ and $MnO_2$ impurities.

FIG. 4: The back scattering image of $La_{0.7}Ce_{0.3}MnO_{3+\delta}$ bulk.   The scattered white droplet and the black blocks with micrometers size are identified to be $CeO_2$ and Mn-O impurities. The main phase, the gray grains, is $La_{0.85}Ce_{0.02}MnO_{3+\delta}$.

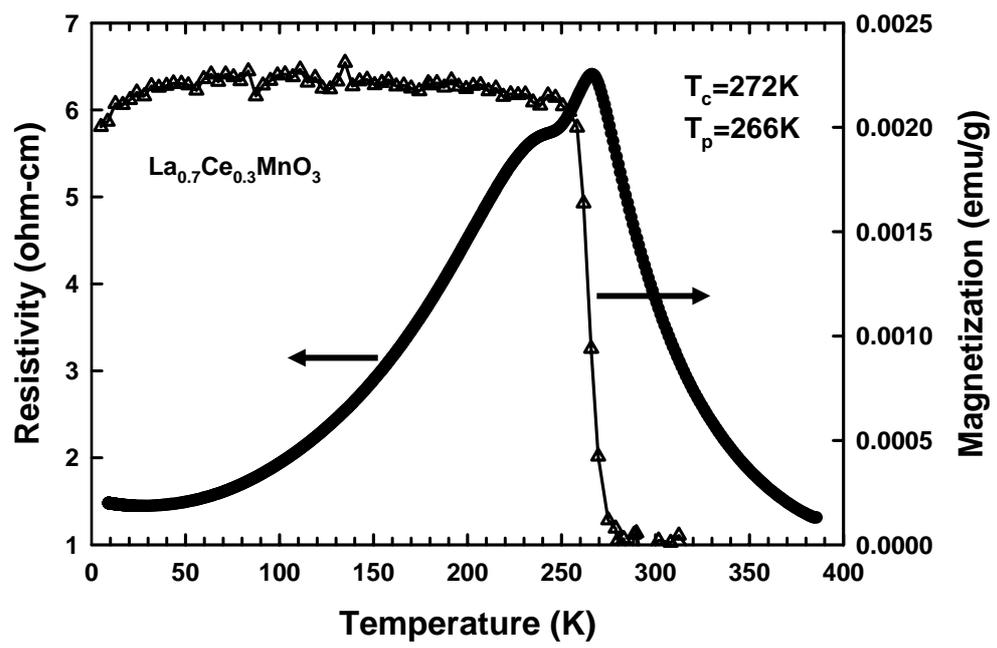

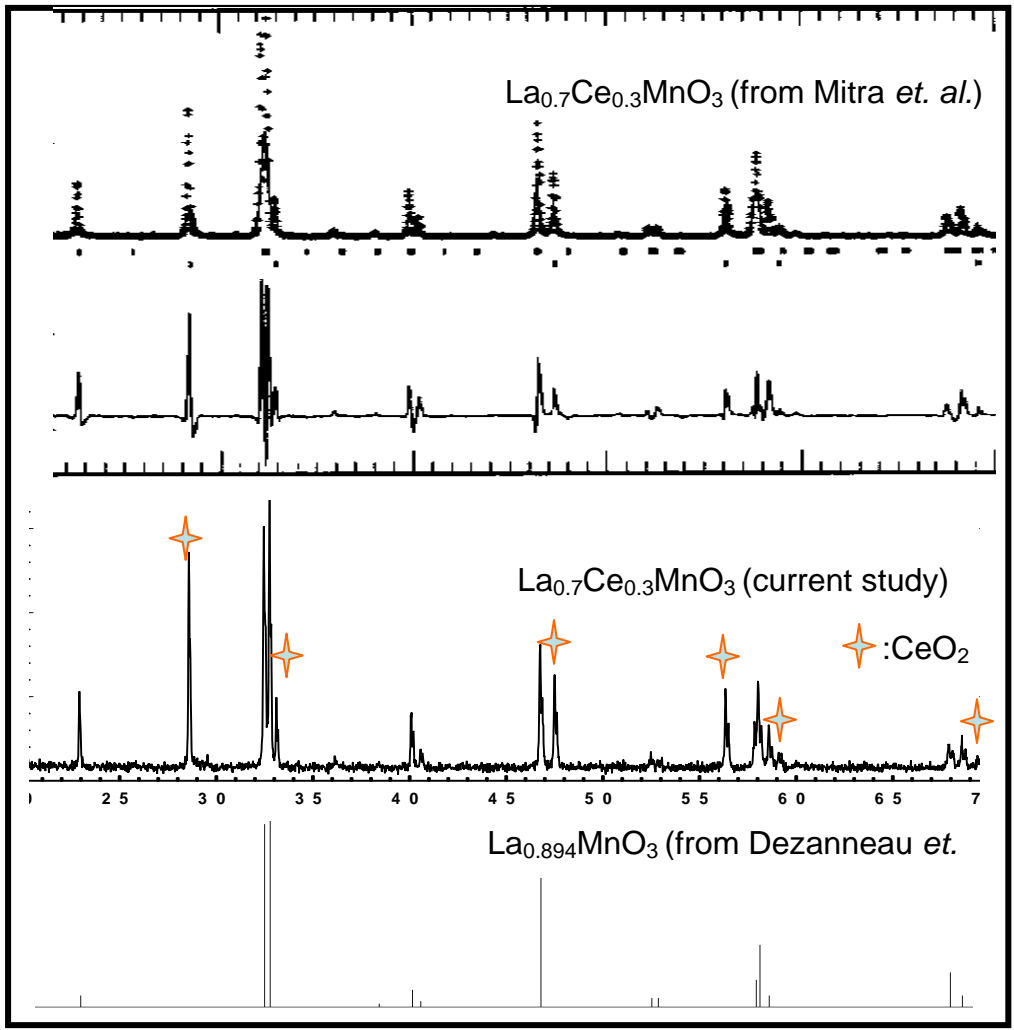

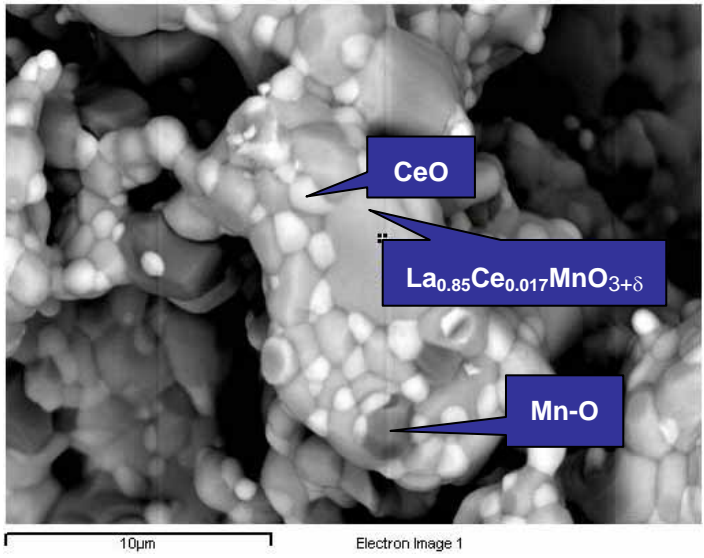

Electron Image 1

10μm

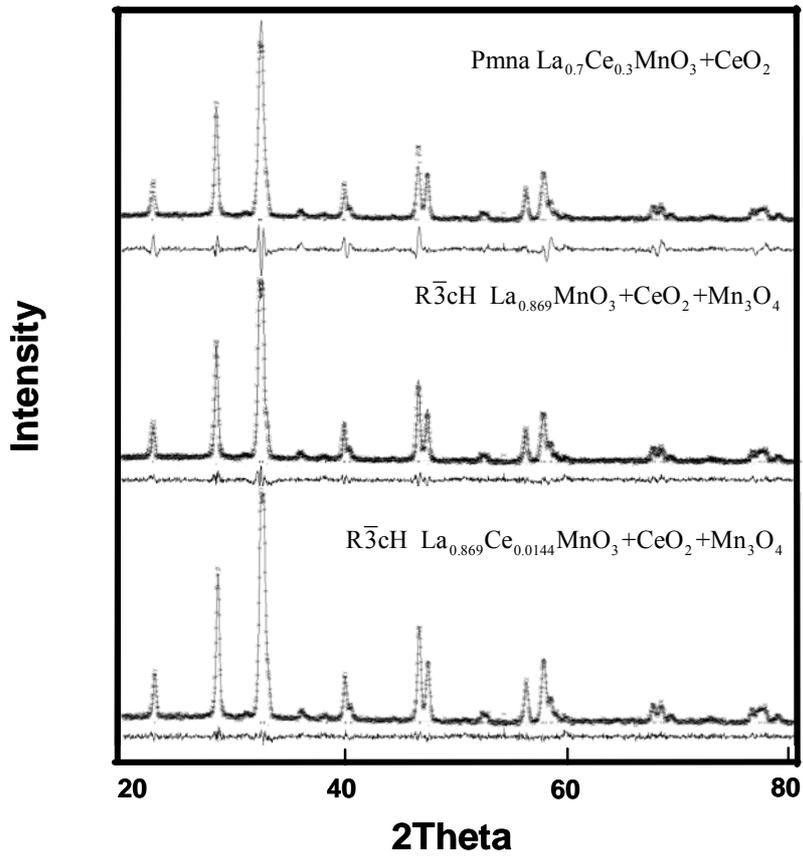